\documentclass[aps,prd,showpacs,nofootinbib,preprintnumbers,twocolumn]{revtex4}

\usepackage{bm}
\usepackage{latexsym}
\usepackage{natbib}
\usepackage{url}
\usepackage{dcolumn}
\usepackage{color}
\usepackage{amsfonts,amssymb,amsmath}
\usepackage{graphicx,epsfig}
\usepackage{psfrag}
\usepackage{subfigure}
\usepackage{natbib}
\usepackage{hyperref}
 
\begin{document}
\title{On counterterms in cosmological perturbation theory}
\author{Gaurav Goswami}
\email{gaurav@iucaa.ernet.in}
\affiliation{IUCAA, Post Bag 4, Ganeshkhind, Pune 411007, India.}
\date{\today}
\begin{abstract}
Cosmological perturbation theory is the theory of fluctuations 
(scalar as well as tensor) around the inflationary cosmological background solution.
It is important to understand the details of the process of renormalization 
in this theory. In more familiar applications of quantum field theory, 
the dependence on the external momenta of the dimensionally regulated 
expression of the one-loop contribution to a correlator determines the 
number of counter terms (and their forms) required to renormalize it.
In this work, it is pointed out that in cosmological perturbation theory, 
though this still happens, it happens in a completely different way such that
in the late time limit, the information about the number and forms of 
counter terms required gets erased.
This is to be compared with what happens in spontaneous symmetry breaking where 
the use of fluctuation fields around a chosen vacuum seems to suggest that more 
counter terms shall be needed to renormalize the theory than are actually required.
We also comment on how the field strength of curvature perturbation, $\zeta$, 
could get renormalized.
\end{abstract}
\maketitle

\section{Introduction}

The methods of quantum field theory (QFT) are some of the most universally applicable techniques in all of physics.
The same techniques apply to (among other things) elementary particle physics \cite{quantum}, 
to statistical and condensed matter physics \cite{tsvelik2007quantum}, 
to the non-linear theory of structure formation in the universe \cite{2002PhR...367....1B,2013arXiv1301.7182P} , 
to the theory of turbulence in fluids and plasmas \cite{Zhou20101}, 
to the calculation of spectrum of gravitational waves from inspiralling binaries \cite{2006IJMPD..15.2293G}, 
to quantum optics \cite{gerry2005introductory},
to black hole thermodynamics \cite{2010arXiv1011.5875K} as well as
to the calculation of correlations of primordial metric perturbations in inflationary cosmology \cite{2009arXiv0907.5424B}.
In this light, it is not surprising that the details of calculations in these varied scenarios,
though similar, are not exactly identical. Thus, it is important to find out which ideas 
apply universally to all these problems and which ones are specific to the applications we 
are most familiar with (e.g. scattering problems). To aid the discussion, in this work, 
the familiar version of QFT shall be referred to as \textquotedblleft usual\textquotedblright~ QFT.

It is a well known fact that inflationary cosmological perturbation theory is almost like a QFT 
of fluctuations around a time-dependent background solution \cite{Maldacena-2003}. 
Consequently, cosmological perturbation theory derives most of its calculational machinery from 
the usual QFT. 
Still, there are many differences, and in this work, we look at some of these differences. 
The interactions of metric perturbations cause corrections to cosmological correlations 
calculated in linear perturbation theory.
This can cause not only primordial non-Gaussianity \cite{Maldacena-2003}, but also, higher 
order corrections to e.g. the two-point function \cite{SZ_all_loops}.

Just like in the usual QFT, in cosmological perturbation theory too, while calculating correlations 
at sufficiently high order, we encounter expressions which are ultraviolet (UV) divergent. In usual QFT, 
we cancel these infinities by (i) collecting the divergent terms with similar dependence on external 
momenta, (ii) finding which terms (allowed by the symmetries of the theory) in the action
can lead to the terms in the correlations with the said dependence on external momenta, and finally, 
(iii) adding these counterterms to the old action (which is now called the renormalized action while 
the sum of the renormalized action and the counterterm is the bare action) 
in order to get correlations which are UV finite at every step of the calculation.

Does the same procedure work in cosmological perturbation theory too? 
As the analysis of this work illustrates, there are subtleties associated with this, one thus has to be 
extremely careful (in this context, see \cite{PhysRevD.72.043514,PhysRevD.74.023508,SZ2010}). 
In cosmological perturbation theory, one is studying fluctuations around a time dependent background 
solution, thus, the action of these fluctuations is not Lorentz invariant. Thus, we can not simply 
list all the possible terms in the action (moreover, since the typical interactions are irrelevant, 
infinite couterterms shall be required). All this ensures that it is difficult to spot the couterterms 
in cosmological perturbation theory. Given this situation, one could ask, given a correlator (e.g. the 
two-point function), which counterterms shall we need to renormalize it at one loop?

From our experience in usual QFT, we are used to spotting the counterterms by looking at the expression 
for dimensionally regulated correlators. This is because the expression for the correlator (with external 
lines amputed) is of the form $\sum_n c_n (k^2)^n$ (where, $k$ is the Lorentzian momentum of the external 
line) i.e., a polynomial in the external momenta with divergent coefficients.  

Unlike in usual QFT, in cosmological perturbation theory, the dimensionally regulated 
logarithmically divergent two-point correlator (for external momentum $k$) is (at late times)
of the form 
\begin{equation}
 \left[ \sum_n c_n (|{\bf k}|)^n \right] \left[\sum_m a_m (-k \eta)^m \right] \; ,
\end{equation}

where, $k$ is the 3-momentum of the external line and $\eta$ is the conformal time at which the 
correlator is evaluated.
We thus have an additional factor involving a polynomial in $-k \eta$:
it is this extra piece which determines the forms and number of counterterms needed to renormalize the theory.
During inflation, we are typically interested in the ${\eta \rightarrow 0^-}$
limit of the correlations. 
If this limit is taken before regularizing the UV divergent momentum
integrals, the information about the forms of counterterms needed and the number of counterterms 
needed to get rid of UV divergences gets erased. Since the number of counterterms also determines 
the number of observations actually needed to fix the renormalized parameters, this is a bad news. 
On the other hand, regularization of momentum integrals at non-zero $\eta$ is a much more 
involved task to perform and hence the results are much less transparent. 
All this illustrates that many properties of the familiar usual QFT calculations do not 
apply to cosmological perturbation theory. Thus, the connection between QFT and cosmological 
perturbation theory is quite subtle and continues to offer surprises.

While calculating the loop corrections to cosmological correlations, it is often assumed that 
the process of renormalization can be performed and the emphasis usually is on the non-trivial 
logarithmic running which may turn up (see, e.g., \cite{PhysRevD.72.043514}). 
E.g. Senatore and Zaldarriaga \cite{SZ2010} have studied renormalization of 
two-point function in cosmological perturbation theory. Their main focus has been on the nature 
of logarithmic running. 
In contrast, in the present work, we look at the actual process of renormalization and the 
associated subtleties. 

We begin in Sec. \ref{sec:usual_qft} by recalling how counterterms are found in usual QFT.
Then, in sec. \ref{sec:cosmo_pert}, after introducing the particular regime of effective field 
theory of inflation for which we present the arguments about the UV divergent two-point correlator, 
we shall describe how the case of cosmological perturbation theory is so different from the usual
QFT in so peculiar a way. We summarize the results in Sec. \ref{sec:summary}. 
We have set $\hbar = c = 1$.

\section{Counterterms in usual QFT:} \label{sec:usual_qft}

In usual QFT, LSZ reduction formula ensures that the most relevant quantity to evaluate is 
the vacuum expectation value of time ordered product of the Heisenberg picture fields.
While evaluating the Fourier transforms of such correlators, at sufficiently high order in perturbation theory 
one encounters UV divergences.
E.g. one could encounter quadratically divergent integrals of the form (for illustrative 
purposes, we work with Euclidean integrals)
\begin{equation}
 I_Q = \int \frac{d^4 \ell}{(\ell^2 + \Delta)} \; ,
\end{equation}
which, on dimensional regularization gives $I_Q = \Delta \left( \frac{\mu^2}{\Delta} \right)^{\delta/2} F(\delta)$,
where, $\delta = 4-d$, $\mu$ is the fake renormalization scale (which inevitably gets 
introduced while performing dimensional regularization) and $F(\delta)$ is a dimensionless 
function of $\delta$ which contains poles of $\delta$. If $F$ has a simple pole and its Laurent series 
expansion of is $F = F_{-1} \delta^{-1} + F_0 + F \delta + \cdots $, then,

\begin{equation} \label{eq:usual_qft_quad}
 I_Q = \Delta \left( \frac{F_{-1}}{\epsilon} + \frac{F_{-1}}{2} \log \left( \frac{\mu^2}{\Delta} \right) + F_0 + \cdots \right) \; .
\end{equation}
Typically, $\Delta$ is a polynomial in the external momentum $k^2$($=-E^2+{\bf p}^2$) or the masses
(and often, the Feynman parameters). 
The presence of $\Delta$ in the above expression causes the dimensionally regulated UV divergent 
integral to be a sum of two parts; the first part is polynomial in the external momentum with divergent 
coefficients and the second one is a finite function of the external momenta, the masses and 
the fake renormalization scale $\mu$. E.g. while evaluating the two-point function in $\phi^4$ 
theory in $d=4$, the corresponding $\Delta$ turns out to be simply $m^2$, so this is a trivial example.
On the other hand, while renormalizing the two-point function for $\phi^3$ in $d=6$, we find that 
the Fourier transform of the two-point function is given by an expression of the form 
\begin{equation} \label{eq:2_pt_usual_QFT}
 {\cal G}(k^2) = {\cal G}_{\rm Free}(k^2) + 
 {\cal G}_{\rm Free}(k^2) \left[ \Pi(k^2) \right]  {\cal G}_{\rm Free}(k^2) + \cdots \; ,
\end{equation}
where ${\cal G}_{\rm Free}(k^2)$ is the Feynman propagator.
$\Pi(k^2)$ is the contribution of the one-loop diagram (with external lines amputated)
and is given by a quadratically divergent integral and the expression for dimensionally 
regulated $\Pi(k^2)$ is of the form 
\begin{eqnarray} \label{Pi_usual_QFT}
\Pi(k^2) = A_2\left( \frac{1}{\epsilon} \right) k^2 + A_0 \left( \frac{1}{\epsilon} \right) m^2 
+ {\cal F}(k^2,m,\mu) \; .
\end{eqnarray}

The following points are important to notice
\begin{enumerate}
 \item $\Pi(k^2)$ has two kinds of contributions: (1) a polynomial in the external momentum $k^2$ 
 (the $m^2$ term can be thought of as the term $(k^2)^0$),
 (2) another function ${\cal F}$, which depends on, among other things, the fake scale $\mu$.
 \item The coefficients of the polynomial in the external momentum $k^2$, the $A_i$, are divergent,
 they are functions of $1/\epsilon$, on the other hand, the function ${\cal F}$ is finite.
 \item The forms of the terms in the polynomial i.e. how they depend on the external momenta, dictate
 the form and number of counter-terms needed to be introduced in the Lagrangian in order to cancel 
 the UV divergences of the theory. E.g. in the above case, they tell us that we need two counter-terms 
 to renormalize the two-point function and they shall be of the form $- \frac{A}{2} (\partial \phi)^2 $ 
 and $- \frac{B}{2} m^2 \phi^2$. 
 In fact, this is how we get convinced that the correlators can be renormalized at all. 
 \item In contrast, for a logarithmically divergent integral 
  \begin{equation}
    I_L = \int \frac{d^4 \ell}{(\ell^2 + \Delta)^2 } \; ,
  \end{equation}
 following the above procedure shall simply give 
  \begin{equation} \label{eq:log_usual}
    I_L = \left( \frac{F_{-1}}{\epsilon} + \frac{F_{-1}}{2} \log \left( \frac{\mu^2}{\Delta} \right) + F_0 + \cdots \right) \; .
  \end{equation}
 It is clear that for a logarithmically divergent integral, the dependence on external momenta 
 does not turn up.
 We could say that for this case, the polynomial is just one. This causes the counterterm to be 
 just a factor such as $Z_g g$ (where $g$ is the coupling constant and $Z_g$ is the renormalizing 
 $Z$ factor). 
 \item The function ${\cal F}$ has this property that in the limit $k^2 \gg m^2$ and $k^2 \gg \mu^2$,
 it takes up the form ${\cal F} \sim k^2 \log \left( \frac{k^2}{\mu^2} \right)$. 
\end{enumerate}

It is important to notice that all this holds good even when we regularize the UV divergent integrals in
some other way.
Similarly, it is not just a property of this theory: it can be easily seen for other theories 
(e.g. Yukawa theory, $\phi^4$ theory, QED): in general, 
the pole part of any sub-divergence-free diagram is a polynomial in its external momenta. 
It has been argued that this holds good also in other theories, where the
residues of the poles always contain the external momenta and masses as low-order polynomials
(see \cite{'tHooft1972189}, \cite{PhysRevD.10.1201},
ch. 9 of \cite{kleinert2001critical}, pg. 148 of \cite{hooft1994under}).
Thus, a polynomial in external momenta plus a function which is 
logarithmic in external momenta is a generic feature of Green's functions in 
\textquotedblleft usual\textquotedblright~QFT.

At this point it is worth re-emphasizing that if we wish to find the number of counter terms needed to 
renormalize a correlator or the form of these counter terms (things we wish to find, even for a 
non-renormalizable theory), the dependence of the Green's function on external momenta is a very 
important tool. 

\section{Counterterms in cosmological perturbation theory} \label{sec:cosmo_pert}

In inflationary cosmological perturbation theory, the most relevant correlator is the 
late time limit of $n-$point function of (Heiseberg picture) comoving curvature perturbation,
$\zeta$ on a constant time hyper-surface
\begin{equation}
\lim_{\eta \rightarrow 0^-} 
\langle \Omega | \zeta(\eta,{\bf x}_1) \zeta(\eta,{\bf x}_2) \cdots \zeta(\eta,{\bf x}_n) | \Omega \rangle \; .
\end{equation}
Homogeneity of the inflationary background implies that all such correlators shall be invariant under translations
in space at a fixed time, which implies that the Fourier transform of the above correlator 
shall be of the form 
\begin{equation} \label{eq:def_G}
 F(\eta, {\bf k}_1, \cdots, {\bf k}_n) = (2 \pi)^{3}~ \delta^3({\bf k}_1 + \cdots + {\bf k}_n)~ 
 \tilde{\cal G}(\eta, {\bf k}_1, \cdots, {\bf k}_n) \; .
\end{equation}
It is worth mentioning that whenever we talk about the usual QFT, we shall be dealing with four dimensional 
Lorentzian momenta while whenever we talk about cosmological perturbation theory, we shall 
be dealing with three dimensional Euclidean momenta.
At sufficiently high order in perturbation theory, one expects to encounter Feynman diagrams with loops. 
This issue, in cosmological perturbation theory has been studied in great detail in the last few years. 
Beginning with \citep{PhysRevD.72.043514}, there was a debate about whether these loop corrections to the 
cosmological correlations shall freeze at late times. Recently (see \cite{SZ_all_loops,2013JHEP...02..151A}), it is claimed 
to be shown that this shall surely happen at all loops. 
The familiar primordial power spectrum $\Delta^2_{\zeta}(k)$  is defined by the Eq.
\begin{equation}
 \lim_{\eta \rightarrow 0^{-}} \langle \zeta(\eta,{\bf k}) \zeta(\eta,{\bf k}') \rangle = 
 \frac{2 \pi^2}{k^3}
~ \delta^3({\bf k} + {\bf k}') ~ \Delta^2_{\zeta}(k) \; ,
\end{equation}
for slow roll inflation with a single (canonical) scalar field, the lowest order (i.e. tree level) contributions 
to power spectrum, assuming Bunch-Davies vacuum (in the limit $\eta \rightarrow 0^{-}$), 
is given by (the classic result)
\begin{equation} \label{eq:PS:free}
  \Delta^2_{\zeta}(k)|_{\rm tree} = \frac{1}{2 \epsilon(\eta_k)} \left(\frac{H(\eta_k)}{2 \pi M_{\rm Pl}} \right)^2 \; ,
\end{equation}
where $\eta_k$ is the conformal time when the mode in question crosses Hubble radius (i.e. when $k = a H$). 
Notice that, on comparing with Eq (\ref{eq:def_G}), it becomes clear that for two-point function, 
\begin{equation}
  \lim_{\eta \rightarrow 0} \tilde{\cal G}(\eta, {\bf k})|_{\rm tree} = 
  \frac{1}{4 \pi k^3} \Delta_{\zeta}^2 (k) = \frac{1}{32 \pi^3 k^3} \left( \frac{H^2}{\epsilon M_{\rm Pl}^2 } \right)\; .
\end{equation}
Since the result in Eq (\ref{eq:PS:free}) is obtained in linear theory, it corresponds to a free theory 
calculation. A well known fact is that Eq (\ref{eq:PS:free}) implies that, the dimensionless Primordial 
Power spectrum is a power law
\begin{equation}
 \Delta_{\zeta, {\rm Free}}^2 (k) = A_s \left( \frac{k}{k_0} \right)^{n_s - 1} \; ,
\end{equation}
where $A_s$ is the spectral index while $n_s$ is the spectral amplitude and $k_0$ is a pivot scale, 
see \cite{2009arXiv0907.5424B}.
This implies that 
\begin{equation} \label{eq:G:zeta:free}
 \tilde{\cal G}_{\rm Free}(k) = \frac{A_s}{4 \pi k_0^3} \left( \frac{k}{k_0} \right)^{n_s - 4} \; .
\end{equation}

The leading interactions of the metric fluctuations are typically due to cubic operators, so, 
it is expected that the loop correction to $\tilde{\cal G}$ shall be of the form (notice 
that the mass dimension of $\tilde{\cal G}$ is $-3$) 
\begin{equation}
 \tilde{\cal G}_{\rm 1-loop} = k^3 (\tilde{\cal G}_{\rm tree})^2 \times ({\rm factors}) \; .
\end{equation}
The factors on the RHS can involve various non-trivial logarithmic runnings e.g. 
$\log a \sim Ht$ or $\log kL$ \cite{2011JCAP...01..023G,2010JCAP...08..006B} 
or $\log (H/\mu)$ \cite{SZ2010}, while a running of the form
$\log k/\mu$ is not possible since this shall not leave $\tilde{\cal G}$ invariant under 
the transformation $x \rightarrow \Lambda x$, $k \rightarrow k/\Lambda$ and 
$a \rightarrow a/\Lambda$ \cite{SZ2010}. 

\subsection{The $\dot{\pi}^3$ theory}

While evaluating correlations for an interacting quantum field  on an accelerating universe, 
one can encounter, apart from the familiar UV divergences, relatively unfamiliar divergences 
too. 
To avoid having to deal with these unfamiliar divergences, and still deal with a realistic 
model of cosmological perturbations which has simple interactions, we work with 
\textquotedblleft the theory of large $\dot{\pi}^3$ interactions \cite{SZ2010}.\textquotedblright~ 
The Effective Field Theory (EFT) \cite{EFT2008} of inflation provides the most general 
framework for systematically studying the dynamics of 
fluctuations around an inflationary background solution. 
The action of the theory of fluctuations can be expanded in powers of the relevant fluctuation 
field (and also in powers of the slow-roll parameters such as the Hubble flow functions). 
The theory is first formulated in unitary slicing of the perturbed spacetime in which $\delta \phi$ 
vanishes (and all the dynamics lies in the metric) and then general covariance is restored by introducing 
the Stueckelberg field (denoted by $\pi$ in the following). 
It turns out that if one chooses to ignore ${\cal O}(\epsilon^2)$ terms in the 
action (which also corresponds to the interaction terms which give rise to primordial non-Gaussianity 
for a canonical scalar field, see \cite{Maldacena-2003}), and one chooses to fix the sound speed of 
fluctuations to unity (i.e. we wish to only consider the cases in which $c_s \rightarrow 1$),
\footnote{Here, the decoupling limit has already been taken and so the terms in the action of EFT 
which cause Stueckelberg field to mix with gravity already vanish, see \cite{EFT2008} for details.}
the leading order interactions for the Stueckelberg field shall be captured by terms of the form
(ignoring ${\cal O}(\pi^5)$ terms, see \cite{EFT2008})
\begin{equation}
 \frac{M_3^4}{6} \left(g^{00} + 1 \right)^3 = - \frac{2}{3} M_3^4 \left[ 2 \dot{\pi}^3 + 3 \dot{\pi}^4 
- \frac{3}{a^2} \dot{\pi}^2 (\partial_i \pi)^2 \right] \; .
\end{equation}
 
It is important to recognize that, thanks to the EFT formulation, it is very easy to identify a 
regime in which the Stueckelberg field has non-negligible self-interactions without violating 
the slow-roll nature of the background solution. Since the symmetry arguments can not fix the 
value (or sign) of the coefficient $M_3$ (which should be determined from observations, see 
\cite{Planck-2013-NG} for the latest limits),
one can write  $M_3^4 (t) = - c_3 (t) M^4$ where $M$ is a mass scale characterizing the 
interaction. An extra shift symmetry can be imposed requiring that the time dependence of 
$c_3$ is negligibly weak.

In this regime, the action of the $\pi$ field becomes (see \cite{SZ2010})
\begin{eqnarray} \label{S_SZ2010}
 S=\int d^4x \; a^3 \left[ -\dot{H} M_{\rm Pl}^2 \left(\dot\pi^2-\frac{1}{a^2}(\partial_i\pi)^2\right) + \right. \\ \nonumber
\left. \frac{2}{3}c_3 M^4\left(2\dot\pi^3+3\dot\pi^4-3\frac{1}{a^2}\dot\pi^2(\partial_i\pi)^2\right)\right]\ ,
\end{eqnarray}

This is perhaps the simplest possible interacting cosmological perturbation theory. 
It is also observationally interesting (see \cite{Planck-2013-NG}). 
This is the most general kind of interactions which are not slow roll suppressed and 
which are leading order when we impose the requirement that $c_s \rightarrow 1$. 
To connect to the usual perturbation variables, one can make a gauge transformation 
to comoving gauge (see \cite{Cheung-consistency}) and find that $\zeta = - H \pi + {\cal O}(\pi^2)$.

We would like to emphasize again that for the inflationary background caused by canonical scalar fields,
the interaction terms in the action of $\zeta$ are ${\cal O}(\epsilon^2)$ so that the interactions
we are dealing with are different from those. Moreover, as is obvious from Eq (\ref{S_SZ2010}), 
in this regime the interaction terms are much simpler (and fewer) as compared to e.g. those in
\cite{Maldacena-2003}. Thus, if we wish to try anything new e.g. 
loop corrections in cosmological perturbation theory and we want a regime which is realistic but 
which is also simple, this \textquotedblleft theory of large $\dot{\pi}^3$ interactions\textquotedblright 
is the best possible choice \cite{SZ2010}.

Given the action (Eq (\ref{S_SZ2010})), the Hamiltonian can be readily 
worked out and then one can use the in-in formalism (see \cite{PhysRevD.72.043514,Contour_rotation} 
and references therein) to evaluate the two-point function at one-loop. 
In the rest of the present work, we shall focus on only the $\dot{\pi}^3$ term in the action 
given by Eq. (\ref{S_SZ2010}). This interaction term leads to two contributions to the two-point 
function only one of which is UV divergent (see \cite{SZ2010}).
On canonical normalization, it becomes apparent that the $\dot{\pi}^3$ interaction 
is of mass dimension $+6$. In the action, a dimension six operator is expected to be accompanied 
with a factor of $1/{\Lambda_U}^2$, where $\Lambda_U$ is the 
\textquotedblleft unitarity bound\textquotedblright~ of the theory.
This is what happens, on canonical normalization, 
$\pi_c \equiv \sqrt{- 2 \dot{H} M_{\rm Pl}^2 } ~ \pi$, and 
the coefficient of $\dot{\pi}_c^3$ operator turns out to be 
\begin{equation}
 \frac{4}{3} \cdot \frac{c_3}{(2 \epsilon)^{3/2}} \cdot \frac{M^4}{H^3 M_{\rm Pl}^3} 
= \frac{4 c_3}{3} \cdot \frac{1}{\Lambda_U^2}
\end{equation}
where $\Lambda_U$ is the energy scale at which this (non-renormalizable) theory becomes strongly coupled
(i.e. the perturbative calculations are valid only at energy scales much smaller than this scale). We 
have,
\begin{equation}
 \frac{1}{{\Lambda_U}^2} = \frac{M^4}{(2 \epsilon)^{3/2} H^3 M_{\rm Pl}^3}
\end{equation}
In correlators, it is expected that the (three-line) interaction vertex of $\dot{\pi}^3$ theory
is always going to be accompanied with factors of the form
$\frac{4 c_3}{3} \frac{H^2}{\Lambda_U^2}$ since $H$ is the energy scale of the inflationary 
\textquotedblleft experiment.\textquotedblright~ 
This suggests that for $\dot{\pi}^3$ theory,  
\begin{equation} \label{gen_exp}
 \tilde{\cal G}_{\rm 1-loop} = k^3 (\tilde{\cal G}_{\rm tree})^2 \times \left( \frac{H^2}{\Lambda_U^2} \right)^2
 ({\rm factors}) \; .
\end{equation}

Dimensional analysis and homogeneity of the background suggest that the correlator 
$\langle\zeta_{\vec k}(\eta)\zeta_{\vec k'}(\eta)\rangle$
shall be of the form 
\begin{equation}
 \langle\zeta_{\vec k}(\eta)\zeta_{\vec k'}(\eta)\rangle = 
 \frac{\delta^3(\vec k + \vec k')}{k^3} \times \left({\rm rest} \right) \; ,
\end{equation}
where the rest terms have to be dimensionless. 
In the one loop calculation, the rest terms shall contain UV divergent momentum integral. 
Thus, apart from a few numerical factors and a factor of
$(c_3^2 M^8/\epsilon^4 M_{\rm Pl}^8)$, the rest terms shall be of the form
\begin{equation} \label{eq:loop_SZ_gen}
 \int d^3{\vec k}_1 d^3{\vec k}_2 \delta^3( {\vec k} + {\vec k}_1 + {\vec k}_2 ) 
 f(\eta, H, {\vec k}, {\vec k}_1, {\vec k}_2) \; .
\end{equation}
Power counting makes it clear that the mass dimension of $f$ is $-3$.
Since this integral is dimensionless,
naively, we'd expect that on dimensional regularization it would give
\begin{equation}
 \left( \frac{k}{\mu} \right)^\delta F(\delta) \; ,
\end{equation}
where $F(\delta)$ is a dimensionless function (which contains poles of $\delta = D-3$).
This result shall not leave $\tilde{\cal G}$ invariant under 
the transformation $x \rightarrow \Lambda x$, $k \rightarrow k/\Lambda$ and 
$a \rightarrow a/\Lambda$. Thus, it is not correct and 
a detailed calculation (by Senatore and Zaldarriaga \cite{SZ2010}) shows that 
in fact the UV divergent momentum integral gives 
\begin{equation}
 \left( \frac{k}{\mu} \right)^\delta F(\delta) G(\delta,k,-\eta, H) \; ,
\end{equation}
and where $G$ is another dimensionless function and when $\eta = 0$, it is of the 
form
\begin{equation}
 G = 1 + \delta \log (-c H \eta_k) + \cdots \; ,
\end{equation}
where $c$ is an ${\cal O}(1)$ constant. 
This changes the logarithmic running to $\log (H/\mu)$ \cite{SZ2010}.

\subsection{Counterterms}

A careful look at the argument presented by Senatore and Zaldarriaga \cite{SZ2010}
(to establish that the logarithmic running is $\log (H/\mu)$) also tells 
that when $\eta = 0$, only one kind of divergence is present. This may suggest 
from the arguments in familiar applications of QFT (Sec. \ref{sec:usual_qft}) 
that we need just one counterterm to cancel the UV divergences in the two-point 
function in this theory. Thus, when $\eta = 0$, we have (see \cite{SZ2010}),
\begin{equation}
 \tilde{\cal G}_{\rm 1-loop} \sim \frac{1}{k^3} 
 \left( \frac{F_{-1}}{\epsilon} + \frac{F_{-1}}{2} \log \left( \frac{H(\eta_k)}{\mu} \right) + F_0 + \cdots \right) \; ,
\end{equation}
where a factor of $(c_3^2 M^8/\epsilon^4 M{\rm Pl}^8)$ is understood to sit in the 
front, apart from some numerical factors. This implies that 
\begin{equation} \label{G_loop_asym}
 \tilde{\cal G}_{\rm 1-loop} \sim k^3 \tilde{\cal G}_{\rm tree}^2 
 \left( \frac{F_{-1}}{\epsilon} + \frac{F_{-1}}{2} \log \left( \frac{H(\eta_k)}{\mu} \right) + F_0 + \cdots \right) \; ,
\end{equation}
with a factor of $(c_3^2 M^8/\epsilon^2 M_{\rm Pl}^4 H^4)$ in the front. This 
should be compared with Eq. (\ref{Pi_usual_QFT}) and (\ref{eq:usual_qft_quad}).

At this point it is worth reminding ourselves that the theory that we are dealing with is a 
non-renormalizable theory so that the countererm (CT) needed is not necessarily one of the 
terms we have already written down in the action Eq. (\ref{S_SZ2010}).
From Eq. (\ref{gen_exp}), it is clear that the loop correction shall have a factor of 
$(H/\Lambda_U)^4$. Again, by dimensional analysis, it is clear that a single vertex of dimension 
$+8$ operator can give this factor. But since we want the CT to renormalize the two-point function, 
it better have two external lines. Thus, counterterm shall be a dimension $+8$ quadratic operator.
On canonical normalization, $\pi_c$ has dimension $+1$ and shift symmetry forbids any polynomials in 
$\pi_c$ to be present in the action. Thus, we can only take derivatives, hence CTs can only be 
operators of the form $(\partial^3 \pi_c)^2$. 
The derivatives that we can take are either w.r.t. time or w.r.t. space,
since we want to write a rotationally invariant action, the only options are
the square of $\partial_t \partial_t \partial_t \pi $, 
the square of $\partial_t \partial_i \partial_i \pi$, 
and $\partial_t \partial_t \partial_i \pi \partial_t \partial_t \partial_i \pi$.
This means that by dimensional analysis, there are three possible candidates for the 
CTs. On the other hand, we have only one \textquotedblleft kind\textquotedblright~ 
of divergent term present in the dimensionally 
regulated expression for the 
$\langle\zeta_{\vec k}(\eta)\zeta_{\vec k'}(\eta)\rangle$ when $\eta = 0$ 
since there is just a monomial of $k^3$ in the front
(see Eq. (\ref{G_loop_asym}) as compared to Eq. (\ref{Pi_usual_QFT})).
In this case, we cannot determine the coefficients of the operators in the CT Lagrangian
in any unique way.

When $\tilde{\cal G}_{\rm 1-loop}$ is worked out for $\eta$ which is non-zero but still
such that $-k\eta \ll 1$, then we get an expression of the form (save for some numerical factors)
\begin{eqnarray}
 \tilde{\cal G}_{\rm 1-loop} &\sim& k^3 ~\tilde{\cal G}_{\rm tree}^2 ~\frac{c_3^2 M^8}{\epsilon^2 M_{\rm Pl}^4 H^4}  \times \\ \nonumber 
 && \left( \frac{F_{-1}}{\epsilon} + \frac{F_{-1}}{2} \log \left( \frac{H(\eta_k)}{\mu} \right) + F_0 + \cdots \right) \times \\ \nonumber
 && \left( \sum_{i=0}^{n} c_i (- k \eta )^{i} \right) \; ,
\end{eqnarray}
i.e., we get an extra term multiplied to the $\eta = 0$ result which is a polynomial, not in the external 
momentum $k$ but in $-k\eta$.
Thus, in this case, even for a dimensionless integral (see the discussion after Eq.(\ref{eq:loop_SZ_gen})) 
we get an additional polynomial, but it is an effect of having non-zero $\eta$. This is to be compared 
with Eq.(\ref{eq:log_usual}) in flat spacetime, where a dimensionless integral which leads to a logarithmic 
divergence gives no polynomials in external momenta and hence the correspondng CTs are trivial.

What is important is the fact that the quadratic CTs (of the form required) also end up giving a polynomial 
in $-k\eta$ and hence renormalization can be performed in this limit. 
But in order to renormalize, we need to determine the (unknown) coefficients of the operators in the CT 
Lagrangian. In the case of familiar field theory (see pt. (4) after Eq. (\ref{Pi_usual_QFT})), each 
individual CT gives a different kind of dependence on the external momenta, on the other hand here, 
that is not the case, we get a polynomial in external momenta whose coefficients are linear combinations 
of the coefficients of the operators in the CT Lagrangian.
Thus, unlike the case of familiar field theory, 
even in the case with $-k\eta \ll 1$ but $\eta$ being non-zero, we cannot just look at the 
expression for dimensionally regulated correlator and read-off the number of CTs required 
or their forms.
Thus, to determine the (unknown) coefficients of the operators in the CT Lagrangian, we need to solve a 
set of linear equations, it so happens that there are four linear Eqs. in the three unknowns and there 
still is a unique solution. Geometrically, this is like having four planes in the three dimensional Euclidean 
space and still they all passing through one common point. The choice of the renormalization conditions shall 
determine this point.
This may raise the following concern: could it be that if we find $\tilde{\cal G}_{\rm 1-loop}$ for $\eta$ such 
that $-k \eta \gtrsim 1$, we end up having many more equations and only three unknowns? Would the solution 
be guaranteed to exist in that case?

But most importantly, in the limit $\eta \rightarrow 0$, only one divergent term is left and thus in taking this 
limit, we end up erasing the information about the form of CTs or their number completely. 
This is similar to what happens in e.g. spontaneous symmetry breaking. If we consider the $Z(2)$ 
symmetric renormalizable scalar field theory 
${\cal L} = - \frac{1}{2} (\partial \phi)^2 - m^2 \phi^2 - \lambda \phi^4/24$.
When $m^2 > 0$, three CTs are enough to absorb all the infinites in the theory. The same is true 
when $m^2 < 0$, but in that case, we can also write the same theory as 
${\cal L} = - \frac{1}{2} (\partial \rho)^2 - \lambda v^2 \rho^2/6 - \lambda v \rho^3/6 - \lambda \rho^4/24$
(where $v = + (6 |m^2|/\lambda)^{1/2}$ and $\rho = \phi - v$) and looking at this Lagrangian, it may appear 
that we shall need more CTs for the $\rho^3$ term as well as to cancel the tadpole (i.e. divergent one-point function)
it shall cause. In this case, the change of variables from $\phi$ to $\rho$ seems to suggest that we shall 
need more CTs while in reality, we do not. In contrast for the case we are dealing with, the process of 
taking $\eta \rightarrow 0$ limit suggests that we shall need fewer CTs, while in reality we do not.

\subsection{Remarks on renormalization}

In $\dot{\pi}^3$ theory, the field is $\pi$ while the various parameters appearing in the action are 
$\epsilon, H/M_{\rm Pl}$ and $c^3 M^4$. The parameter $c^3 M^4$ can be constrained from the observations 
of Primordial Non-Gaussianity in the CMB sky (and it has been constrained by the Planck collaboration
\cite{Planck-2013-NG}).
There also are (upper) limits on the values of $\epsilon$ and $H/M_{\rm Pl}$
\cite{2013arXiv1303.5082P}. If we knew the actual 
values of these parameters (rather than just the limits), we expect that we could work in the 
On-shell (OS) renormalization scheme to perform the actual renormalization of the theory. 
In usual QFT, while renormalizing the two-point function in OS scheme, we choose the counter-terms such that 
(1) the divergent part in the loop integral (the $1/\epsilon$ term) gets canceled,
(2) the dependence on the fake renormalization scale gets cancelled, and,
(3) the rest of the part of the counterterm is chosen such that $\Pi(-m^2) = 0$ and $\Pi'(-m^2) = 0$, 
 see \cite{quantum},
 these conditions ensure that the parameter $m$ appearing in the Lagrangian is the physical mass 
 and the field strength is normalized (thus, at $k^2=-m^2$, the propagator has a simple pole with unit residue).
 In Eq. (\ref{G_loop_asym}), we can use 
\begin{equation}
 \log \left( \frac{H(\eta_k)}{\mu}\right) = 
 \log \left( \frac{H(\eta_k)}{H(\eta_{k_0})}\right) + \log \left( \frac{H(\eta_{k_0})}{\mu}\right) \; ,
\end{equation}
where $k_0$ is a pivot scale to absorb the $\mu$ dependent term in the CT, but instead of the two conditions 
(given by Lehmann-K\"all\'en\ form of the exact propagator), in the usual flat spacetime field theory, 
we shall need three (since we have three CTs) and we have no equivalent of the Lehmann-K\"all\'en\ spectral  
representation of the propagator.
Thus, in cosmological perturbation theory, we seem to have no straightforward way to apply the OS scheme since 
we can not naturally relate the measured correlations to the physical values of the parameters of the 
Lagrangian of the theory. At this stage, we can again go back to usual field theory to seek inspiration about how 
to perform renormalization.
In usual QFT, in any chosen renormalization scheme, when we have the expression for e.g. ${\cal G}(k^2)$ 
(the notation of Eq. (\ref{eq:2_pt_usual_QFT})), we find $M$ such that at $k^2 = - M^2$, ${\cal G}(k^2)$ has
a simple pole, we realize that $M$ is the physical mass (as opposed to the parameter which merely turns up
in the renormalized part of the Lagrangian). At $k^2 = - M^2$, if the residue of ${\cal G}(k^2)$ is $R$
(and in general, it is not unity), we define the renormalized field as $\phi_{\rm ren} \equiv \phi/{\sqrt R}$, 
so that the residue of the pole for renormalized field is unity.

In cosmological perturbation theory, we could proceed in the following way: if loop corrected power 
spectrum is
\begin{equation}
 \Delta^2_{\zeta} (k) = A_s \left( \frac{k}{k_0} \right)^{n_s - 1} \left( 1 + g(k,\mu,c_3 M^4, H) \right) \; ,
\end{equation}
then, in general at $k = k_0$, $\Delta^2_{\zeta} (k) \neq A_s$. But if $\Delta^2_{\zeta} (k_0) = R(\mu)$, then,
let us redefine $\zeta$ such that 
\begin{equation}
 \zeta_{\rm ren} (\mu) \equiv \sqrt{ \frac{A_s}{R(\mu)} } \zeta \; ,
\end{equation}
and this is how field strength renormalization could be done in cosmological perturbation theory. Moreover, 
one could use the observations of $A_s$, $n_s$ and $dn_s/d \log k$ (the running of the spectral index) 
to fix the finite parts of the three CTs of this theory. At this stage however, the observational 
constraints on most parameters: $M_3$ \cite{Planck-2013-NG}, $\epsilon$, $H$ (during inflation), 
$dn_s/d \log k$ \cite{2013arXiv1303.5082P} are not good enough to perform this procedure.
Notice that had we taken the $\eta \rightarrow 0$ limit before renormalizing, we could not have known that 
we need three CTs and the above would not have been possible.

\section{Summary} \label{sec:summary}

In this work, we explored issues of renormalization in cosmological perturbation theory.
In the more familiar applications of QFT, a logarithmically divergent loop integral has 
a trivial polynomial dependence on external momenta e.g. 
\begin{equation}
 {\cal I}_1(k) = \int_0^{\infty} \frac{d \ell}{\ell + k} = C - \log k \; ,
\end{equation}
(where $C$ is divergent), on the other hand, for a quadratically divergent integral ${\cal I}_2$, one gets,
\begin{equation}
 {\cal I}_2(k) = \int_0^{\infty} \frac{\ell d \ell}{\ell + k} = a + b k + k \log k \; ,
\end{equation}
where $a$ and $b$ are divergent. 
In general, in usual QFT, every diagram with external lines amputated and with no sub-divergences is of 
the form
\begin{equation}
 G = \sum_{i = 0}^{n} A_i (1/\epsilon) (k^2)^i + G_{\log} (k^2,m,\mu) \; ,
\end{equation}
(with $k^2$ being Lorentz invariant) 
where, for the case of a logarithmically divergent diagram, 
only the the $i=0$ term is present. 
In cosmological perturbation theory, in contrast, even for a logarithmically
divergent diagram, one gets, when $-k\eta \ll 1$,
\begin{equation}
 G =  \left[ A_0 (1/\epsilon) k^3 + G_{\log} (k,m,\mu) \right] \left[ \sum_{i=0}^{n} c_i (- k \eta )^{i} \right] \; .
\end{equation}
We argued that if one intends to perform renormalization, one needs to identify the CTs from the
expression of dimensionally regulated correlators. Unlike the case of usual QFT, here, when
$-k\eta \ll 1$, the expression for dimensionally regulated correlator is a polynomial in 
$-k\eta$, and taking the limit $\eta \rightarrow 0$ erases information about the forms and number 
of CTs required to renormalize the theory.
We thus realized that one should be very careful in taking the limit in which the external time 
$\eta$ goes to zero.
We also explored how the process of renormalization could be performed in cosmological
perturbation theory (e.g. how $\zeta$ shall undergo field strength renormalization). 
This illustrates the many subtleties and surprises associated with field theoretic aspects of 
cosmological perturbation theory.

 \noindent {\bf Acknowledgment:} 
 The author thanks T. Padmanabhan (IUCAA, Pune), U. Yajnik (IIT-Bombay, Mumbai) 
 and R. Rangarajan (PRL, Ahmedabad) for comments on a previous draft of the manuscript.

\label{Bibliography}
\bibliographystyle{h-physrev3}  
\bibliography{Bibliography_QFT,biblio_inf_tmp,QFT-ubiquity}  

\end{document}